\documentclass[a4paper]{article}
\usepackage{graphicx}
\usepackage[latin1]{inputenc}
\usepackage{times}

\graphicspath{{./}{../images/}}

\begin{document}

\title{Impact of Obstacles on the Degree of Mobile Ad Hoc Connection Graphs}

\author{Cédric Gaël Aboue-Nze, Frédéric Guinand and Yoann
  Pigné\thanks{Authors are alphabetically ordered. This work was partially supported by ANR SARAH project. Yoann Pigné is funded by French Ministry of Research. Cédric Aboue Nze is funded by the Ministry of Research of the Gabon Republic.}\\
University of Le Havre\\
LITIS EA 4108\\
BP 540, 76058 Le Havre - FRANCE\\
}  
\date{}

\maketitle

\begin{abstract}
What is the impact of obstacles on the graphs of connections between stations in Mobile Ad hoc Networks? In order to answer, at least partially, this question, the first step is to define both an environment with obstacles and a mobility model for the stations in such an environment. The present paper focuses on a new way of considering the mobility within environments with obstacles, while keeping the core ideas of the well-known Random WayPoint mobility model (a.k.a RWP). Based on a mesh-partitioning of the space, we propose a new model called RSP-O-G for which we compute the spatial distribution of stations and analyse how the presence of obstacles impacts this distribution compared to the distribution when no obstacles are present. Coupled with a simple model of radio propagation, and according to the density of stations in the environment, we study the mean degree of the connection graphs corresponding to such mobile ad hoc networks. 

\end{abstract}

\section{Introduction\label{sec:intro}}

Mobile ad hoc networks (MANET) are self-organizing telecommunication networks formed by moving devices able to communicate with each other in a decentralized way, without relying on any infrastructure. Such devices are also called stations or nodes. In this document, we will keep the term station for naming the mobile devices within the network and the term node to name the vertices of the corresponding connection graphs.
The core of MANET functioning resides in the communications between stations or more precisely in the possibility for the stations to communicate. 
From the modelling point of view, graphs are probably the best candidates for representing such systems. Within such graphs, vertices are associated to stations and an edge links two vertices if and only if their corresponding stations can communicate, 
in the context of this paper this occurs when the stations are in communication range with each other. 
Of course, the performances of algorithms operating on such networks strongly depend on graphs properties and on their evolution. The study of these properties may be conducted either by simulation or by probabilistic analyses.  
In the latter case, metrics like distribution of degrees, mean number of connected components or mean path length require for their examination: the knowledge of the spatial station distribution within the environment; the conditions of the existence of links between stations according to the signal propagation model; the stations properties and the environment characteristics. In the present paper, we focus on a probabilistic analysis of the spatial node distribution when the mobility of stations is constrained coupled with the coverage constrained by the obstacles.

From our point of view, a mobile ad hoc network (a MANET or a DT-MANET) may be defined as a 4-uplet $(D,S,E,B)$ where:

\begin{itemize}
\item $D$ is a set of devices with their own properties (communication range, speed limits, power, etc),
\item $S$ is a signal propagation model,
\item $E$ is an environment defined by a set of characteristics (speed limits, building/obstacles characteristics, forbidden areas, predefined paths, etc), and,
\item $B$ corresponds to the station behavior for moving within the environment. This behavior may include the choice of the destination and the strategy for choosing the path from the current position to the destination point. 
\end{itemize}

This paper is not the first, and certainly not the last one, to deal with spatial node distribution, however, as far as we know, the analyses performed so far were conducted only for unobstructed environments, namely environments free of obstacles and considering the well-known Random WayPoint mobility model (RWP).
In that work, we determine the spatial node distribution for environments containing obstacles. To that end, we begin with a careful analysis of some existing mobility models taking into account obstacles and we propose to consider a mobility scheme that keeps the core ideas of the classical Random WayPoint mobility model but that is readily adaptable to various environments, including partially obstructed ones. This new model is denoted RSP-O-G, an acronym for summing up the mobility behavior, the main characteristics of the environment and the constraints on the authorized paths. RSP-O-G stands for {\em Random Shortest Path - Obstacles - Grid}. It means that stations move within an environment represented by a grid and containing obstacles. Between a source $s$ and a destination $d$, a station moves along a shortest path in the grid between $s$ and $d$. As many shortest paths may exist between two nodes in a grid, the choice of the path is randomly performed by the station. Using this model we are able to compute the spatial node distribution in an environment that contains obstacles. In addition, at the end of this paper, we discuss the possibility for this model to represent close-to real life situations.  

The document is organized in three parts. In Section \ref{sec:rspo}, both the environment and the mobility scheme are described and compared with existing mobility models that take into consideration obstacles. Section \ref{sec:meandeg} is devoted to the description of the method for computing locally the mean value of the degree. This Section begins with the method for computing the spatial node distribution for the newly defined mobility scheme, and exposes the kind of computations needed for estimating the station coverage in presence of obstacles. Section \ref{sec:experiments} presents some experiments. Section \ref{sec:extensions} concludes temporarily this work and discusses some extensions.

\section{Random-Based Shortest Path Mobility Models with Obstacles \label{sec:rspo}}



In the litterature, the notion of mobility model usually encompasses mobility behavior of stations, environment characteristics, signal propagation model but also constraints about pathways.
 

In the context of Mobile Ad Hoc Networks, Bai and his colleagues in\cite{Bai2004} proposed a classification of mobility models in four main categories based on some specific characteristics: Random-based models, models with Temporal Dependencies when movements are based on the history of previous moves, models with Spatial Dependencies when movements depend on other nodes moves, and models with Geographic Restrictions when movements of nodes are limited by obstacles or constrained by environmental characteristics (highways or streets for instance). 

In our case, the proposed model belongs both to the Random category since stations behavior is random-based and to the Geographic Restrictions category since the environment described may contain some obstacles. We think however that these criteria (randomness, geographic restrictions, spatial dependencies and temporal dependencies) are not disjuntive but additive. Indeed, as we will see in Section \ref{sec:stateoftheartmmo}, the majority of mobility models belonging to the Geographic Restriction category are also based on random choices. Thus, we consider that studying the mobility of stations in MANET requires: a) the definition of the environment characteristics, its size and shape, the presence of obstacles, its heterogeneity/homogeneity, etc; b) the definition of the stations mobility behavior, including the rules for choosing the next destination, decision that may depend on other nodes or on the history of previous moves and the strategy used for moving from the current position to the destination; and c) a set of constraints limiting or defining the authorized paths within the environment. These three elements are sufficient for our goal: computing the spatial distribution of stations and their local coverage. 

Given that point of view, we examine previous works dealing with mobility of stations within environments partially obstructed or within environments defining authorized paths that restrain the mobility of the stations. 
%
%

\subsection{Mobility Models with Obstacles \label{sec:stateoftheartmmo}}



In the context of MANET, the mobility model that may be considered as the reference is the well-known Random WayPoint mobility model (RWP) since it serves as a basis for many mobility schemes. 

In that model, the environment is bounded and obstacle-free, such that the mobility of the stations is not constrained. The behavior of stations for moving is simple. Each station randomly chooses a destination point located anywhere in the environment, selects a speed that belongs to a speed interval and moves from its current position to the destination point in straight line. Then, the station may stay for a while at the same place before moving again according to the same process. We can qualify this mobility behavior as a Random destination Shortest Path moving within a bounded and unconstrained environment since within an obstacle-free environment the straight line is the shortest path. 

Previous works that are identified to be closely-related with our study are those that deal with mobility models considering constrained paths. This includes models based on obstructed environments or based on unobstructed environments but considering constrained paths. We begin with the former category and in particular with the article of Jardosh and his colleagues \cite{Jardosh2005}. 

In this work, the authors propose a combinaison of both a new mobility and a new signal propagation models. In our work we are only interested in the mobility and environment characteristics but not in the signal propagation model. The environment described in their work allows the placement of obstacles. From the set of obstacles, they build a Voronoi tessellation from which a set of constrained pathways is deduced. Stations can only move along these paths, but they can pass through obstacles. Initially, stations are randomly distributed over the pathways. The mobility behavior of stations is based on a random choice of the destination and the shortest path for reaching that place is chosen. 

Several other works dealing with mobility in environments with obstacles are based on this article.  
Based on the remark that the set of constrained paths obtained from the Voronoi tessellation is rather limited, Huang in \cite{Huang2005} proposes another way of defining constrained paths around obstacles using a double Delaunay triangulation of the space. This method shortens the mean values of the shortest paths within the environment, but the number of distinct shortest paths remains limited. Another more recent extension of Jardosh's work is due to Babaei, Fathi and Romoozi. While in the work of Jardosh \cite{Jardosh2005}, the stations choose their destination randomly anywhere on the paths defined from the Voronoi tessellation, in \cite{Babaei2007}, the authors propose to add Hot-Spots to the environment, getting closer to a urban-like environment. 

While the presence of obstacles in the environment impacts node mobility, it is possible to obtain the same results on mobility without considering obstacles but by defining constrained paths. The Manhattan and the Freeway mobility models defined in \cite{Bai2003} belong to this category. Both models express the fact that in urban spaces the pathways are constrained. In the Freeway scheme, the constrained paths are made of several parallel paths: the lanes. Each station can only move on one lane. In the Manhattan model, the pathways correspond to a set of horizontal and vertical lanes. The station may change lane at each intersection according to a probability.

Finally, among the set of works taking into account mobility schemes with obstructed environments or with constrained pathways, some consider group-based mobility scenarii, corresponding to the Spatial Dependencies category in \cite{Bai2004}. For instance, in \cite{Kristoffersson2005}, while the environment is the same as in \cite{Jardosh2005}, the stations are divided in two types: reference and secondary. Each secondary node is attached to a reference node. Reference nodes are only moving on the constrained paths. Secondary nodes may move anywhere in the environment but their movements are constrained by the reference node they are attached to. Following a similar idea, Williams and Huang \cite{Williams2006} propose a mobility model based on a Boids-like principle. 
While very interesting and considering obstructed environments, these works are out of scope of the present study.

Let us now describe in detail our {\em Random Shortest Path - Obstacle - Grid} (RSP-O-G) model. 

%

\subsection{The RSP-O-G model \label{sec:rspomodel}}


The original question we asked was: how can it be possible to extend the well-known Random Waypoint Mobility Model when some obstacles are present within the environment, while limiting the constraints on paths like it is done in previous works?

We attempt to answer this question by analyzing station behavior in RWP mobility model, and we conclude that RWP is based on two core ideas.  
First, stations choose their destination randomly. Second, stations are choosing the shortest path for moving toward their destination.  As it was underlined in the previous Section, most of existing mobility models with obstacles also rely on these two ideas, however, the set of possible shortest paths between a source and a destination is strongly limited by the imposed constraints. Keeping that in mind we propose to transpose this principle, random choice of the destination and shortest path moving by considering mesh-based environments with obtacles.  

We define RSP-O-G as the coupling of a mobility behavior and an environment containing obstacles but in which stations may only move into the grid. Between a source and a destination point, all shortest paths are supposed to be equiprobable. We consider an environment with obstacles that may be randomly distributed. Obstacles are simply defined by the removing of nodes and edges in the mesh representing the environment. These obstacles may be buildings or other infrastructures that may be passed through niether by the stations nor by the signal. 
Station behavior is twofold. It concerns both the destination choice and the strategy for moving from the current position to the destination. As previously mentionned, the choice of destination is random and the strategy for moving from one place to the destination consists in choosing one shortest path.



\section{Mean Degree \label{sec:meandeg}}



For this theoretical analysis, the stations are considered identical: they all have the same radio transmission range ($r$). We also suppose the environment to be a square or more generally a rectangle, defined by its length $L$ and its width $W$. The density of stations is defined by the mean number of station by surface unit is denoted by $\rho$.


The question we address in this papier is: {\em how can we compute the mean degree of the vertices in an environment containing obstacles?} 

If we consider an obstacle-free environment and a uniform distribution of stations in that environment ($\rho$ stations/surface unit). If we assume the coverage of the station $C(S)$ to be a disk which radius corresponds to the radio transmission range of the station, then, the mean value of the degree of such stations is equal to $\pi r^2 \times \rho$. However, if the density of stations is not uniform on the whole environment, and if the coverage of the station may vary according to its position in the environment, then the mean value of the degree of a station located in $(x,y)$ may be estimated to: $\mbox{deg}(S_{(x,y)}) = C(S_{x,y}) \times \rho_{(x,y)}$.

Then, for computing the mean value of the degree locally in every node of the mesh, we have to both: estimate the coverage of station at each position, and compute the probability of presence of one station at each node of this environment based on the mobility scheme defined by the RSP-O-G model. This will constitute the matter of Subsections \ref{sec:snd} and \ref{sec:coverage}.

\subsection{Spatial Node Distribution Analysis \label{sec:snd}}

In addition to the works dealing with mobility models taking obstacles into account, there exist a second corpus of articles that are closely related to our work: the ones that deal with spatial node distribution in the context of mobile ad hoc networks. The spatial distribution of stations determines the probability for a station to be located at a given place. Any station going its way in the mesh is also making a path in the graph. This problem has been extensively studied by Bettstetter and his colleagues \cite{Bettstetter2003,Bettstetter_Wagner_2002}, but mainly in the context of the basic RWP mobility model. The main hypotheses for the spatial distribution computation are i) start and destination positions of a node are independently and identically distributed over the environment, and ii) the ergodic hypothesis may be applied to RWP. In our case the environment is a mesh, the location is a node in the graph induced by the mesh. Any station going its way in the mesh is also making a path in the graph. 

In the RWP mobility model, stations choose destinations and then move straight ahead to these places. Stations are naturally going the shortest paths. In the graph, if the measure of the shortest path is proportionnal to the number of edges of the path, then there may be many shortest paths between two nodes. This number $N_{sp}$ in a grid, between $s$ and $t$ is equal to: $N_{sp}(s,t) = \frac{\displaystyle (|x_s - x_t|+|y_s-y_t|)!}{|x_s - x_t|! \times |y_s - y_t|!}$
where $x_s$ and $y_s$ are the Cartesian coordinates of the source node $s$ and $x_t$ and $y_t$ are the coordinates of the destination $t$ (the target node).  


In general, the probability of presence of one station on a place between the source node $s$ and the destination $t$ is proportional to the length of the path. In a grid, the probability of presence of a station on a node of the grid is also proportional to the length of the shortest path but not only. Since there may exist many shortest paths, the presence probability is also proportional to the number of shortest paths that go through the given node. Thus, the probability for one station to be on the node $n$ when going from node $s$ to $t$ is:
 
\begin{equation}
P_{sp}^{(n)}(s,t) = \frac{\displaystyle N_{sp}^{(n)}(s,t)}{\displaystyle N_{sp}(s,t)} . L_{sp}(s,t)^{-1}  
\label{eq:distribution_P_n} 
\end{equation}

where $N_{sp}^{(n)}(s,t)$ is the number of shortest paths between $s$ and $t$ that go through node $n$. $N_{sp}(s,t)$ is the number of shortest paths between $s$ and $t$ and $L_{sp}(s,t)$ is the length of the path.   

The difficulty in this model is the computation of $N_{sp}^{(n)}(s,t)$. Note that within a simple grid without forbidden area, $N_{sp}^{(n)}(s,t) = N_{sp}(s,n) \times N_{sp}(n,t)$. In all other cases, this value can be determined thanks to Dijkstra's algorithm  which gives a shortest path tree (a digraph actually), from a source node to all the destinations in the graph. Once Dijkstra's algorithm applied, we get a directed acyclic graph ($DAG$) from the target $t$ to the source node $s$ ($DAG_{sp}(t,s)$). Then two searches in this $DAG$ give the sought number of paths through $n$. The algorithm consists in 2 labellings, one for each search.  

In the first search nodes and edges are tagged  ($tag1$) with integer values. Nodes are tagged  with the sum of the tags of their incoming edges. One node can only be tagged  if all of its incoming edges are tagged  as well. The tag for any node $n$ is $tag1(n) = \sum_{e \in IN(n)} tag1(e)$
with $IN(n)$ the list of incoming edges of node $n$ in $DAG_{sp}(t,s)$ and $tag1(n)$ the tag of $n$. Edges are tagged  with the tag of their source node ($tag1(e) =  tag1(S(e))$).   
$S(e)$ is the source node of edge $e$ and $tag1(e)$ is the tag of $e$. The first search starts with the target node $t$ tagged with $1$. 

The second search goes the other way in $DAG_{sp}(t,s)$, back from $s$ to $t$ and tags nodes edges with a second tag ($tag2$). Nodes are tagged with the sum of the tags ($tag2$) of their outgoing edges :
$tag2(n) = \sum_{e \in OUT(n)} tag2(e)$.
$OUT(n)$ is the list of outgoing edges of node $n$, and  $tag2(n)$ is the second search tag for $n$. One node can only be computed if all of its outgoing edges are tagged with $tag2$. Edges are tagged  according to their target node, proportionally to their first tag and the node's tags:  

\begin{equation}
tag2(e) = \frac{tag1(e)}{tag1(T(e))} . tag2(T(e)
\label{eq:tag2_e}
\end{equation}

With $T(e)$ the target node of edge $e$ and $tag2(e)$ the second search tag for edge $e$. 

The general algorithm for computing the number of paths that go through each node and edge of the graph for a given couple $(s,t)$ is: 1) compute the shortest path from source node $s$ according to Dijkstra's algorithm and extract the directed acyclic graph $DAG_{sp}(t,s)$ that leads from $t$ to $s$; 2) compute a breadth-first search from $t$ to $s$ in $DAG_{sp}(t,s)$. For each node $n$ in the search:  $i$) compute $tag1(n)$  with the sum of the tags of their incoming edges; $ii$) for each edge in the outgoing edges of $n$, compute $tag1(e)$ with the tag of their source node ($tag1(e) =  tag1(S(e))$); 3) compute a second breadth-first search back from $s$ to $t$ and for each node $n$:  $i$) compute $tag2(n)$ with the sum of the tags ($tag2$) of their outgoing edges; $ii$) For each edge $e$ in the incoming nodes of $n$, compute $tag2(e)$ according to (\ref{eq:tag2_e}).

Then, for each node (and edge) that belongs to the shortest paths between $s$ and $t$, $tag2$ contains the number of such paths that go through this node (or edge). A node (or edge) in the graph with no value $tag2$ has zero shortest path from $s$ to $t$. Finally, the probability of presence on each node and edge can be computed according to (\ref{eq:distribution_P_n}). 

From the hypotheses we said that source and destination positions of a station are indepedently and identically distributed over the environment, thus, the probability of presence is computed for all couples of nodes in the grid. 

%

The time complexity of the algorithm is quite important since it requires many searches in the grid to be computed. Let $n$ be the number of nodes and $m$ the number of edges. For each node in the graph, Dijkstra's algorithm is performed. Then the double search is made from the given node to all the nodes that have not been computed yet. In the worst case, the double search costs $2.m$, $(O(m))$ in time, and is computed $n-1$ times. The complexity of Dijkstra's algorithm is $O(m)$, so the overall time complexity is thus $n(m+mn)$, that is $O(n^2m)$. The space complexity is almost linear. 


\subsection{Coverage in Presence of Obstacles \label{sec:coverage}}

\begin{figure}[!t] 
\centering
\includegraphics[width=0.75\linewidth]{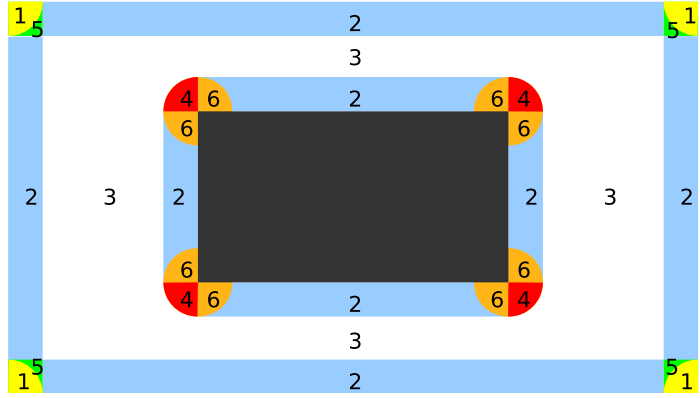} 
\caption{The six zones in which the coverage of the station varies.} 
\label{fig:zones} 
\end{figure}

In order to obtain an estimation of the mean degree of the connection graph, we need, additionaly to the density distribution of presence of station, their coverage, depending on their position with respect to the position of the obstacles. For computing the stations coverage, we need a signal propagation model. We have chosen a simple model for demonstrating the applicability of our approach. The considered signal propagation model is Line-Of-Sight, meaning that an edge exists between two stations if and only if a) they are in communication range with each other, condition fulfilled if their euclidian distance is lower than the minimum value of their coverage radius, and b) if no obstacle is present between them, that is if the stations see each other. Neither diffraction, nor reflection, nor refraction, nor absorption is considered. Perturbations with the emission or with the reception are not taken into account. The signal neither circumvent nor cross the obstacles. The coverage of a station represents the surface covered by the signal of the station. We suppose that the surface covered by a station in a free space is a disk. However, the surface covered by a station is reduced when it is close to either a border of an obstacle. The closer a station to an obstacle or to a border, the smaller its coverage surface. We determine six kinds of zones (see Figure \ref{fig:zones}) for coverage calculations for a station. Indeed, we consider that the distance between the borders and any obstacles and between obstacles themselves is greater than or equal to twice the value of the radio transmission range of any station. 

\begin{figure}[!t] 
\centering
\includegraphics[width=0.7\linewidth]{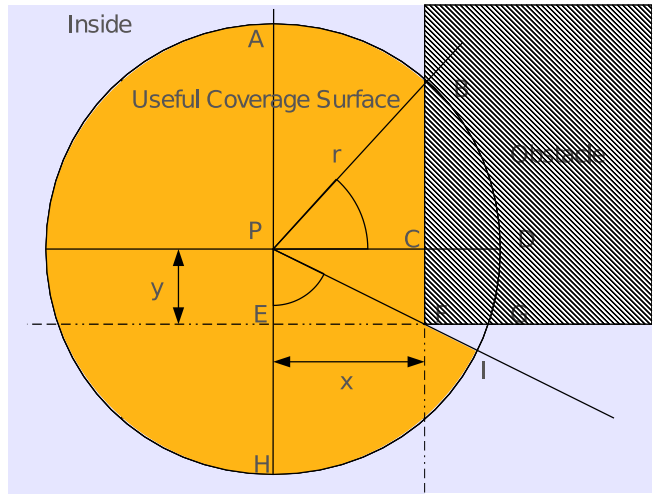} 
\caption{Coverage surface in Zone 6} 
\label{figure3} 
\end{figure}

Here is an example of such a computation for the zone labelled 6 in Figure \ref{figure3}. The total surface covered by the station is equal to half a disk added to the colored surface of PABC added to the triangle PCF added to the surface of the PIH sector. 

The surface of the area delimited by ABCP is equal to a quarter of disk minus the surface of the area delimited by BCD incluced in the obstacle. This area may be deduced from the knowledge of the surface of both the sector PBD and the triangle PBC. Note that $\cos(\alpha) = \frac{x}{r}$ i.e. $\alpha = \arccos(\frac{x}{r})$, then the surface of the sector PBD is equal to $r^2.\arccos(\frac{x}{r})/2$, and the surface of the triangle PBC is equal to $x.r.sin(\arccos(\frac{x}{r}))/2$ given that $BC = r.\sin(\alpha) = r.\sin(\arccos(\frac{x}{r}))$. The surface of BCD is then equal to $r^2.\arccos(\frac{x}{r})/2 - r.x.sin(\arccos(\frac{x}{r}))/2$ and thus, the surface of the area delimited by ABCP is qual to $\pi.r^2/4 + x.r.sin(\arccos(\frac{x}{r}))/2 - r^2.\arccos(\frac{x}{r})/2$. Given that $\tan(\beta) = \frac{x}{y}$, and thus $\beta = \arctan(\frac{x}{y})$, the surface of the sector PIH is equal to $r^2.\arctan(\frac{x}{y})/2$, and the surface of the triangle PCF is equal to $\frac{x.y}{2}$.


Finally, it comes that the coverage surface of a station located in Zone 6 is equal to:
$3.\pi.r^2/4 + x.y/2$ $+ r^2.\arctan(\frac{x}{y})/2$
$+ x.r.\sin(\arccos(\frac{x}{r}))/2$ $- r^2.\arccos(\frac{x}{r})/2$\\
if $(P_x \leq F_x)$ $\&$ $(P_y \geq F_y)$ and,\\
$3.\pi.r^2/4$ $+\ x.y/2$ $+\ r^2.\arctan(\frac{y}{x})/2$
$+ y.r.\sin(\arccos(\frac{y}{r}))/2$ $- r^2.\arccos(\frac{y}{r})/2$\\
if $(P_x \geq F_x)$ $\&$ $(P_y \leq F_y)$.

\section{Experiments \label{sec:experiments}}


This part focuses on the description of an example rather than on the presentation of results of extensive experiments. Indeed, the goal of the current document is to present the general approach for computing characteristics of dynamic graphs extracted from DT-MANET, and to propose a model for that purpose. For illustrating the whole approach, we develop a full example. 

\begin{figure}[!t] 
\includegraphics[width=9cm]{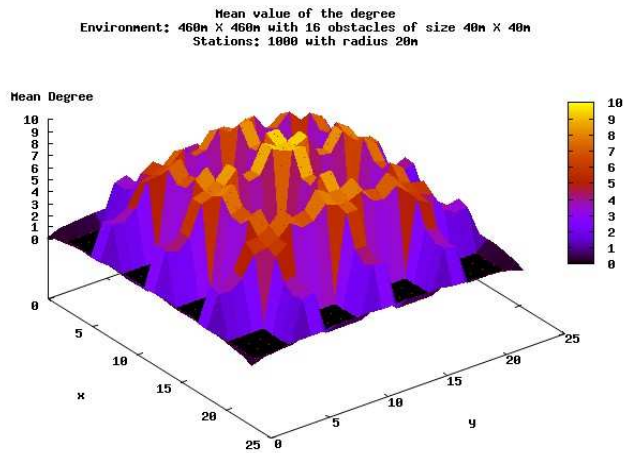} 
\caption{Mean value of the degree.} 
\label{fig:meandegree} 
\end{figure}

The environment is a square of 460$\times$460 meters that contains sixteen square obstacles of the same size (40$\times$40 meters), uniformly distributed over the environment. We have considered a set of 1000 stations with a radio transmission range fixed to 20m. The probability of presence was computed according to the method exposed in Section \ref{sec:meandeg}, and the coverage was computed using the six zones illustrated on Figure \ref{fig:zones}. Figure \ref{fig:meandegree} is the result of the computation of both the probability of presence of the station in each point and the value of the coverage of the station in these points.

\section{Conclusion and Future Works \label{sec:extensions}}


The main goal of this paper was to measure the impact of obstacles on the mean degree within connection graphs stemmed from Delay-Tolerant as well as classical Mobile Ad Hoc Networks while keeping a RWP-like mobility scheme. For that, we propose in a first part a model, called RSP-O-G, for representing both the environment and the mobility of stations in a convenient way for allowing mixed algorithmic/analytic analyses. The computational method relies on the definition of a mesh upon the environment in order to include forbidden areas (corresponding to the obstacles). Within this mesh-based environment, the stations behave according to the rules of the Random WayPoint mobility model: destinations are randomly chosen and the strategy for moving from the current position to the destination is the shortest path.


We have shown that the shape of the spatial node distribution computed by Bettstetter and his colleagues can be recovered using our method, moreover this method enables the computation of such distribution for environments containing obstacles, whatever their shape, size and position, which constitutes a novelty with respect to state-of-the-art. 


Furthermore, the approach, mesh-based environments with forbidden areas, may be extended to many other scenarii in order to build scenarii close to real-life situations. For instance in a urban environment, speed limits associated to the lanes may be different, such that taking longer geographical pathways may be shorter in time. We can simulate such situation in the model by associating to each edge a speed value. The computation remains the same since Dijkstra algorithm remains valid under these conditions. Finally, in relation with the works of \cite{Babaei2007} and with the notion of Hot-Spot, it is possible to define hotspots in our RSP-O-G model by introducing a biais in the probability of choosing sources and/or destinations for the computations, but these two extensions have not yet been implemented.

\bibliographystyle{alpha}
\bibliography{sdc}

\end{document}